\documentstyle[12pt]{article}
\def\al{\alpha}

\def\si{\sigma}

\def\ph{\phi}

\def\om{\omega}

\def\De{\Delta}

\def\fr#1#2{{{#1} \over {#2}}}

\def\vev#1{\langle {#1}\rangle}

\def\frac#1#2{{\textstyle{{#1}\over {#2}}}}

\def\lsim{\mathrel{\rlap{\lower4pt\hbox{\hskip1pt$\sim$}}
    \raise1pt\hbox{$<$}}}
\def\gsim{\mathrel{\rlap{\lower4pt\hbox{\hskip1pt$\sim$}}
    \raise1pt\hbox{$>$}}}
\def\sqr#1#2{{\vcenter{\vbox{\hrule height.#2pt
         \hbox{\vrule width.#2pt height#1pt \kern#1pt
         \vrule width.#2pt}
         \hrule height.#2pt}}}}

\newcommand{\beq}{\begin{equation}}
\newcommand{\eeq}{\end{equation}}
\newcommand{\bea}{\begin{eqnarray}}
\newcommand{\eea}{\end{eqnarray}}
\newcommand{\rf}[1]{(\ref{#1})}

\renewenvironment{thebibliography}[1]
 { \rm
   \begin{list}{\arabic{enumi}.}
    {\usecounter{enumi} \setlength{\parsep}{0pt}
     \setlength{\itemsep}{3pt} \settowidth{\labelwidth}{#1.}
     \sloppy
    }}{\end{list}}

\begin{document}
\titlepage

\begin{flushright}
{COLBY-95-05\\}
{IUHET 308\\}
{July 1995\\}
\end{flushright}
\vglue 1cm

\begin{center}
{{\bf THE EVOLUTION AND REVIVAL STRUCTURE\\
	OF LOCALIZED QUANTUM WAVE PACKETS\\}
\vglue 1.0cm
{Robert Bluhm$^a$, V. Alan Kosteleck\'y$^b$,
and James A. Porter$^a$\footnote{Address after
September 1, 1995: Physics Department,
Cornell University, Ithaca, NY 14853.}\\}
\bigskip
{\it $^a$Physics Department\\}
\medskip
{\it Colby College\\}
\medskip
{\it Waterville, ME 04901, U.S.A\\}
\bigskip
{\it $^b$Physics Department\\}
\medskip
{\it Indiana University\\}
\medskip
{\it Bloomington, IN 47405, U.S.A.\\}

}
\vglue 0.8cm

\end{center}

{\rightskip=3pc\leftskip=3pc\noindent
Localized quantum wave packets can be produced in a
variety of physical systems and are the subject of
much current research in atomic, molecular, chemical,
and condensed-matter physics.
They are particularly well suited for studying the
classical limit of a quantum-mechanical system.
The motion of a localized quantum wave packet
initially follows the corresponding classical motion.
However,
in most cases the quantum wave packet spreads and
undergoes a series of collapses and revivals.
We present a generic treatment of wave-packet
evolution,
and we provide conditions under which
various types of revivals occur in ideal form.
The discussion is at a level appropriate
for an advanced undergraduate
or first-year graduate course in quantum mechanics.
Explicit examples of different types
of revival structure are provided,
and physical applications are discussed.

}

\vskip 1truein
\centerline{\it Accepted for publication in
the American Journal of Physics}

\vfill
\newpage

\baselineskip=20pt

{\bf\noindent I. INTRODUCTION}
\vglue 0.4cm

The advent of short-pulsed lasers has made it possible to
produce and detect coherent superpositions of
quantum-mechanical electron states for a variety of
physical systems.
Once produced,
such superpositions can result in the
formation of localized electron wave packets.
The evolution and dynamics of these wave packets are the subject
of much current investigation in many areas
of physics and chemistry.

One of the most studied applications of wave packets,
both theoretically and experimentally,
is in atomic physics using Rydberg atoms.
Excitation by a short laser pulse produces
a Rydberg wave packet,
i.e.,
a superposition of highly excited single-electron states
\cite{ps,az}.
One aim of investigations on Rydberg wave packets is
to probe the interface between classical and quantum mechanics
\cite{sciam}.
This is feasible because,
when a Rydberg wave packet first forms,
its motion is periodic with the same classical period $T_{\rm cl}$
as a charged particle in a Coulomb field
\cite{tenWolde,yeazell1}.
However,
this motion lasts for only a few cycles,
whereupon quantum-interference effects cause the
wave packet first to collapse and then to undergo a
sequence of revivals
\cite{ps,az,ap,nau1,peres,detunings,sr,sr2}.

The revivals of Rydberg wave packets are characterized
by the recombination of the collapsed wave packet at
a time $t_{\rm rev}$ into a form close to the original shape
that again oscillates with period $T_{\rm cl}$.
This may be viewed in part as a manifestation of the quantum
recurrence theorem
\cite{bocch,almost},
which follows because the time evolution
is driven by discrete energy eigenstates.
In addition to the full revivals,
fractional revivals occur at times that are rational fractions of
the revival time $t_{\rm rev}$ and correspond to the
formation of macroscopically distinct subsidiary wave packets
\cite{ap}.
These have periodicity equal to a rational fraction
of the period $T_{\rm cl}$.
Both full and fractional revivals have been detected in
pump-probe time-delayed photoionization experiments
\cite{yeazell2,yeazell3,meacher,wals}.
More recently,
it has been shown that, for times beyond the revival
time $t_{\rm rev}$,
a new sequence of full and fractional revivals commences
\cite{detunings,sr,sr2}.
These are characterized by a longer time scale
called the superrevival time $t_{\rm sr}$.
At times that are certain rational fractions of
$t_{\rm sr}$,
distinct subsidiary waves again form,
but with a periodicity that is a rational
fraction of the revival time $t_{\rm rev}$.
These long-term fractional revivals culminate with the
formation of a single localized wave packet,
called a superrevival
\cite{detunings},
that more closely
resembles the initial wave packet than does the full
revival at time $t_{\rm rev}$.

The study of revivals is not limited to
Rydberg wave packets in atomic physics.
Wave-packet dynamics is being explored in several areas in physics
and chemistry,
and the structure of revivals is relevant to a variety
of problems.
We list here some examples.
In molecular physics,
a new realm of phenomena involving wave packets has
opened up with the emergence of femtosecond pulse technology
\cite{garraway}.
Ultrashort pulses are used to prepare and probe molecular
wave packets in ways that permit the charting of
chemical processes.
A related application attempts to control chemical
reactions using optimized light pulses
\cite{tannor,warren}.
For example,
frequency-chirped laser pulses have been synthesized to
control the evolution of vibrational wave packets of iodine
\cite{wilson}.
Wave packets have also been produced in semiconductor
quantum-well systems
\cite{leo}.
Their dynamics is similar to that of atomic systems,
allowing for the observation of quantum beats and revivals
\cite{gobel,bs}.
Revival structure has also been observed experimentally
in the evolution of the atomic inversion in
a micromaser
\cite{rwk}.
Some recent theoretical predictions include
fractional revivals for the Jaynes-Cummings model
\cite{avjc},
for wave packets in a Morse-like potential
\cite{vet},
and for an atom bouncing vertically in a gravitational cavity
\cite{cm}.

Several different types of revival structure
can be found in these systems.
Most of them share the feature of exhibiting an
initial periodicity,
often corresponding to that of a related classical problem,
followed by a sequence of collapses and subsequent revivals.
However,
the times at which the revivals occur,
the extent to which they resemble the initial wave packet,
and the question of whether fractional revivals
and/or superrevivals occur
all depend on the quantum system being considered.

In this paper,
we show that the revival structure
of wave packets for different quantum systems
can be treated in a generic fashion
and that classes of quantum systems exist
for which the revival structures are perfect.
We define time scales that govern
the behavior and evolution of a quantum wave packet
and use them to introduce the
various types of revival that can occur.
The results are illustrated with examples
of simple quantum systems,
including the harmonic oscillator,
the infinite square well,
the rigid rotator,
and Rydberg wave packets.
In addition to being standard solvable systems in
quantum mechanics,
these examples also serve as the foundation for
much of the theoretical analysis performed
on actual physical systems of interest in atomic, molecular,
chemical, and condensed-matter physics.
Other examples could be developed via a straightforward
extension of the methods presented here.

The level of treatment and the examples we consider
make this topic appropriate for an advanced undergraduate
or first-year graduate course in quantum mechanics.
The superposition of eigenstates that comprises
a wave packet and the analysis of its time evolution
can be studied either using standard software packages
for symbolic manipulation and graphics
or via computer programs comparable in difficulty to others
successfully employed by undergraduates
\cite{goldberg,picture}.
The treatment of Rydberg atoms
presented here is particularly useful
in discussions of the classical-quantum correspondence
and in much current research in laser-based atomic physics.

This paper is organized as follows.
In the next section,
we describe the generic features of localized quantum
wave packets,
define the time scales that control their evolution,
and discuss the revival structure.
Specific examples are considered in
Sec.\ III.
The autocorrelation of a time-evolved wave function
with the initial wave function provides a striking
illustration of different types of revival structure.
This is described in Sec.\ IV,
along with illustrative examples.
We summarize in Sec.\ V.
Throughout this paper,
we work mostly in atomic units for which
$\hbar = e = m_e = 1$.
The conversions to standard units are
$1~{\rm a.u.} \simeq 5.29 \times 10^{-11}$ m
for lengths and
$1~{\rm a.u.} \simeq 2.42 \times 10^{-17}$ sec
for times.

\vglue 0.6cm
{\bf\noindent II. REVIVAL STRUCTURE OF QUANTUM WAVE PACKETS }
\vglue 0.4cm

{\bf\noindent A. Generic Features}
\vglue 0.4cm

The time-dependent wave function
for a localized quantum wave packet
formed as a one-dimensional superposition of energy eigenstates
may be written as
\beq
\Psi ({\vec r},t) = \sum_{n} c_n
\psi_n ({\vec r}) \exp \left[ -i E_n t \right]
\quad .
\label{expand}
\eeq
Here,
the vector $\vec r$ denotes the position of
the wave function in one or more dimensions.
The label $n$ represents the quantum number
relevant for the sum in the superposition.
The wave functions $\psi_n (\vec r)$ represent
(possibly degenerate)
energy eigenstates with corresponding eigenvalues $E_n$.
Any quantum numbers other than $n$ are suppressed
for convenience.
The coefficients $c_n$ are the weighting factors in
the superposition,
given in terms of the initial wave function by
$c_{n} =
\left< \psi_{n}({\vec r}) \vert \Psi ({\vec r},0) \right>$.

In the general case
the superposition \rf{expand}
may include continuum states,
for which the index $n$ would become continuous
and the sum would become an integral.
However,
our primary interest here is in superpositions of bound states
with negligible continuum contributions,
as these are the type produced in typical experiments
on revivals in wave packets.
Therefore,
in what follows we assume the index $n$ is discrete.
Also,
we assume that the expansion \rf{expand}
is strongly weighted around a mean value $\bar n$ for
the quantum number $n$.
This assumption is reasonable for situations
involving the production of a localized wave packet
using a short-pulsed laser.
The uncertainty principle applied to the laser light
states that
$\tau\De E \ge 1/2$,
where $\tau$ is the duration of the laser pulse and
$\De E$ is the spread in energy.
Exciting a quantum system from its ground state with
a short laser pulse produces a superposition of
states centered on a value $\bar n$ that depends on the
mean frequency of the laser light and has a spread $\si$
in $n$ that is inversely related to the duration $\tau$
of the laser pulse.
In situations of interest $\bar n$ is typically large,
so we assume this in what follows.

We model the weighting probablities $|c_n|^2$
as a gaussian distribution
\beq
\bigm| c_n \bigm|^2 = \fr 1 {\sqrt{2 \pi \si^2}}
e^{- \fr {(n - {\bar n})^2} {2 \si^2}}
\quad .
\label{gauss}
\eeq
This simple assumption provides
a symmetrical distribution in $n$
with mean $\bar n$ and standard deviation $\si$.
It therefore can be used to match the mean value
and spread of states produced by a laser,
though the shapes of the distributions may differ.
The assumption of a gaussian distribution
is unimportant in what follows,
serving primarily to simplify computation
of the wave-packet evolution.
Other weighting distributions
could also be considered.
For example,
coherent states for the simple harmonic oscillator
\cite{klauder}
may be produced by weighting the energy eigenstates
with the coefficients
\beq
c_n = e^{- \fr 1 2 | \al |^2}
\fr {\al^n} {\sqrt{n!}}
\quad ,
\label{cscn}
\eeq
which depend on a complex parameter $\al$.
This distribution is asymmetric in $n$ and is
weighted around $\bar n \approx | \al |^2$
with spread $\si \approx | \al |$.
These coherent states have minimum uncertainty,
exhibit motion with the classical period,
and maintain their shape with time.
Minimum-uncertainty coherent states may also be
obtained for other problems
\cite{nieto},
although the associated wave packets
subsequently collapse and undergo revivals.

The wave packets produced in experiments using
single short laser pulses to excite Rydberg atoms
are localized only in the radial coordinate and have
p-state angular distributions.
They are called radial Rydberg wave packets.
Minimum-uncertainty solutions
have been obtained for these wave packets
\cite{rss,susywave}.
They are examples of squeezed states,
which are characterized by having a shape and
uncertainty product that oscillate with time
\cite{primers}.
To produce Rydberg wave packets that move on an elliptical
trajectory in a plane,
external fields must be applied to mix the angular states.
One scheme for producing elliptical wave packets has been proposed
\cite{gns}.
A squeezed-state description of planar elliptical wave packets
has recently been obtained
\cite{2dss}.
The expressions for the corresponding weighting
coefficients for the radial and elliptical squeezed states
are given in
refs.\ \cite{susywave,2dss}.

\vglue 0.6cm
{\bf\noindent B. Time Scales}
\vglue 0.4cm

The assumption that the weighting probabilities
$| c_n |^2$ are strongly centered around a mean value
$\bar n$ means that only those states with energies
$E_n$ near the value $E_{\bar n}$ enter appreciably into
the sum in Eq.\ \rf{expand}.
This permits an expansion of the energy in a Taylor series in $n$
around the centrally excited value $\bar n$:
\beq
E_n \simeq E_{\bar n} + E_{\bar n}^\prime (n - {\bar n})
+ \fr 1 2 E_{\bar n}^{\prime\prime} (n - {\bar n})^2
+ \fr 1 6 E_{\bar n}^{\prime\prime\prime} (n - {\bar n})^3
+ \cdots
\quad ,
\label{Taylor}
\eeq
where each prime on $E_{\bar n}$ denotes a derivative.

The derivative terms in
Eq.\ \rf{Taylor}
define distinct time scales that depend on $\bar n$:
\beq
T_{\rm cl} = \fr {2 \pi} { |E_{\bar n}^\prime |}
\quad , \qquad
t_{\rm rev} = \fr {2 \pi} {\fr 1 2 | E_{\bar n}^{\prime\prime} |}
\quad , \qquad
t_{\rm sr} = \fr {2 \pi} {\fr 1 6 | E_{\bar n}^{\prime\prime\prime} |}
\quad .
\label{Tcl}
\eeq
The first time scale,
$T_{\rm cl}$,
is called the classical period.
The second time scale,
$t_{\rm rev}$,
is the revival time.
The third time scale,
$t_{\rm sr}$,
is the superrevival time.
In practice,
for all cases of interest
the expansion \rf{Taylor} is such that
$T_{\rm cl} \ll t_{\rm rev} \ll t_{\rm sr}$.

Keeping terms through third order and disregarding an
overall time-dependent phase,
we may rewrite
Eq.\ \rf{expand} as
\beq
\Psi ({\vec r},t) = \sum_{n} c_n
\psi_n ({\vec r}) \exp \left[ -2 \pi i
\left( \fr {(n - {\bar n})t} {T_{\rm cl}}
+  \fr {(n - {\bar n})^2 t} {t_{\rm rev}}
+ \fr {(n - {\bar n})^3 t} {t_{\rm sr}} \right) \right]
\quad .
\label{psi3rd}
\eeq
Since both $n$ and $\bar n$ are assumed to be integers,
the powers of $(n - {\bar n})$ are all integer-valued as well.

A more general treatment of the energies and laser excitation
would allow for noninteger values $n^\ast$ of the
quantum number as well as a laser detuning
with $E_{\bar n}$ off resonance.
The consequences of these additional effects have been
considered for excitations of
Rydberg wave packets in alkali-metal atoms,
which have energies
$E_{n^\ast}$ given by a Rydberg series depending
on quantum defects
\cite{detunings}.
The quantum defects shift the values of $n^\ast$ to
noninteger values and thereby model the effects of the
core electrons in an alkali-metal atom.
Ref.\ \cite{detunings} shows that the effects of quantum
defects on the long-term evolution and revival structure
of a Rydberg wave packet are different from the effects
caused by a laser detuning.
The noninteger values for $n^\ast$ and the laser detuning
cause shifts in the revival time scales and periodicities
that can be distinguished from each other.
The treatment of the wave-packet evolution including these
effects is similar to that given here.
Since the time scales are much simpler for the case
of integer-valued quantum numbers and with the laser
on resonance,
we restrict our analysis here to these cases.

\vglue 0.6cm
{\bf\noindent C. Revival Structure}
\vglue 0.4cm

The expansion \rf{psi3rd}
shows that the time dependence
of the wave function is governed by the three time scales in
Eq.\ \rf{Tcl},
which in turn are controlled by
the dependence of the energy on the quantum number $n$.
In what follows,
we first summarize
the behavior of a wave packet for the generic case
where all three time scales are finite.
We then discuss special classes of quantum systems
for which wave-packet evolution is particularly simple.

For small values of $t$,
the first term in the phase in Eq.\ \rf{psi3rd} dominates.
Evidently,
during this interval motion of the
wave function $\Psi(\vec r,t)$
is approximately periodic in time with period $T_{\rm cl}$.

As $t$ increases and becomes appreciable
compared to $t_{\rm rev}$,
the second term in the phase modulates this behavior,
causing the wave packet to spread and collapse.
However,
at later times near $t_{\rm rev}$,
the second term in the phase in
Eq.\ \rf{psi3rd}
is approximately equal to $2 \pi i$,
which makes it irrelevant,
and once again the motion is governed by the first term.
As a result,
the wave packet regains its initial shape,
and its motion is periodic with the period $T_{\rm cl}$.
This is called a full revival.
For times greater than $t_{\rm rev}$
but much smaller than $t_{\rm sr}$,
the revival cycle repeats.
At every multiple of $t_{\rm rev}$,
the second term in the phase again equals $2 \pi i$
and a full revival occurs.

At special times
that are rational fractions of $t_{\rm rev}$,
the wave packet gathers into a series of subsidiary waves
called fractional revivals.
The motion of these fractional revivals is periodic
\cite{ap}
with period given by
a rational fraction of $T_{\rm cl}$.

The presence of the third-order term
in the time-dependent phase
modulates the whole cycle of periodicity,
collapse, and fractional/full revivals
\cite{sr2}.
At times appreciable compared to the superrevival time scale
$t_{\rm sr}$,
the full-revival structure collapses
and a new system of revivals appears
with features controlled by $t_{\rm sr}$.
At times given by the rational fractions
$\fr 1 q t_{\rm sr}$,
with $q$ an integer multiple of 3,
the wave packet again separates into a series of
subsidiary waves called fractional superrevivals.
In this case,
however,
the motion of the wave packet is periodic with a period
equal to $\fr 3 q t_{\rm rev}$.
Moreover,
at the time $\fr 1 6 t_{\rm sr}$,
the wave function reforms into a single wave packet that
resembles the initial one better than does the full
revival at time $t_{\rm rev}$.
This new structure is called a superrevival.

We have kept only the first three derivative terms
in the energy expansion \rf{Taylor}
and hence have only three time scales in
Eq.\ \rf{psi3rd}.
In principle,
higher-order terms contribute
and lead to a modulation of the full/fractional
superrevivals for times $t \ge t_{\rm sr}$.
In most physical systems,
these higher-order effects are irrelevant.
For example,
for Rydberg wave packets the fourth-order time
scale is comparable to the lifetime of the Rydberg
states and hence the Rydberg wave packets
spontaneously decay before this time scale comes into play.
The superrevival time scale $t_{\rm sr}$,
however,
is several orders of magnitude less than the spontaneous
lifetime as well as the lifetime for blackbody-induced
transitions,
so it is feasible for experiments to detect superrevivals
of Rydberg wave packets.

We can use the above analysis to
identify classes of quantum systems that clearly exhibit
each of the above behaviors.
The simplest nontrivial class consists of
systems which are perfectly periodic and
for which wave packets undergo neither
revivals nor superrevivals.
By perfect periodicity,
we mean that the motion of the wave packets between times
$t=0$ and $t = T_{\rm cl}$ exacly matches the motion
between times $t = mT_{\rm cl}$ and $t=(m+1)T_{\rm cl}$
for any positive integer $m$.
For such systems,
there are no revivals or superrevivals in the sense that
the initial periodic motion is maintained indefinitely.
In this case,
the time scales
$t_{\rm sr}$ and $t_{\rm rev}$ are infinite,
or equivalently $d^kE_n/dn^k \vert_{\bar n} = 0$
for all $k\ge 2$.
This means that wave packets in quantum systems with
energy eigenstates of the form
$E_n = A + B n$,
with $A$ and $B$ independent of $n$ and $B\ne 0$,
exhibit perfect periodic motion and regain their
initial shape at times that are multiples of $T_{\rm cl}$.
The wave packets never fully collapse and there are no revivals
or superrevivals.
Examples of systems of this type
are provided by the
$D$-dimensional harmonic oscillators,
for which the eigenenergies are
$E_n = (n + D/2) \om$,
where $\om$ is the frequency
and $n$ is the principal quantum number.
Since these systems have no higher-order modulating
terms in the time-dependent phase of Eq.\ \rf{psi3rd},
the shape of the wave packet remains unchanged
after each classical period.

The next simplest class consists of systems
for which wave packets undergo revivals
but not superrevivals.
This means
$t_{\rm sr}$ is infinite,
or equivalently $d^kE_n/dn^k \vert_{\bar n} = 0$
for all $k\ge 3$.
In this class are quantum systems
with eigenenergies having $n$ dependence
of the form
$E_n = A + B n + Cn^2$,
where $A$, $B$ and $C$ are independent of $n$
and $C$ is nonzero.
The associated wave packets
undergo motion with the classical period
modulated by the revival phase,
as expected.
However,
in these systems the full revival is a perfect replica
of the original packet,
since the revival structure is not modulated
by a superrevival timescale.
The infinite square well and the rigid rotator
are examples of quantum systems
having energy spectra in this class.

One might suspect the existence of a class
of quantum systems for which the superrevivals
are perfect in the sense that there are no higher-order
contributions to the time-dependent phase
of Eq.\ \rf{psi3rd}.
This would be the case if there existed a
quantum system with bound-state energies having
a cubic dependence on $n$.
However,
the bound-state energies
in nonrelativistic systems with confining potentials
can rise no faster than the power $n^2$
in the limit of large $n$
\cite{ns}.
In principle,
a quantum system might be identified
with energies varying as the cube of $n$
for a restricted range of $n$.
If a wave packet in such a system were constructed
solely as a superposition of energies in this range,
it would have perfect superrevivals.
We are unaware of any physical quantum system
with this feature.
In the remainder of this paper
where superrevivals are considered,
we take the hydrogen atom with $E_n = -1/2n^2$
as our canonical example of a quantum system
with eigenenergies non-quadratic in $n$.
Wave packets in this system exhibit the full range
of behavior described above.

\vglue 0.6cm
{\bf\noindent III. EXAMPLES}
\vglue 0.4cm

In this section,
we illustrate the discussion in Sec.\ II
with examples of wave packets for several different
quantum systems:
the harmonic oscillator,
the infinite square well,
the rigid rotator,
and the hydrogen atom.

\vglue 0.6cm
{\bf\noindent A. Simple Harmonic Oscillator}
\vglue 0.4cm

According to the discussion
in Sec.\ II,
wave packets for the simple harmonic oscillator
never collapse and have no revivals.
We illustrate this by plotting the wave packet in
Eq.\ \rf{expand}
using the gaussian distribution for the $c_n$
coefficients given in
Eq.\ \rf{gauss}
and the standard energy eigenfunctions
$\psi_n(x)$ for the simple harmonic oscillator.
Since the sum in
Eq.\ \rf{expand}
is strongly peaked on the value $\bar n$,
we may truncate the sum after a finite number of terms
and numerically evaluate the result.
For convenience,
we set $\om = 1$ in atomic units.

Figure 1 shows various stages in an orbital period
of a wave packet with
$\bar n = 15$ and $\si = 1.5$.
The wave packet is initially located
at the right-hand turning point of
the corresponding classical oscillator.
Note that,
although the wave packet is localized,
its shape is \it not \rm gaussian
and so it is \it not \rm a coherent state.
The shape therefore changes during the motion.
Nonetheless,
the packet follows the classical motion
with period $T_{\rm cl} = 2 \pi/E_{\bar n}^\prime = 2 \pi$.
The motion is perfectly periodic,
so that each subsidiary figure shows the wave packet at
the given time up to an integer multiple of $T_{\rm cl}$.
The wave packet does not collapse and there are
no revivals or superrevivals.

\vglue 0.6cm
{\bf\noindent B. Infinite Square Well}
\vglue 0.4cm

The energy eigenstates for a particle in an infinite
square-well potential with $0 \le x \le L$
are $\psi_n(x) = \sqrt{\fr 2 L} \sin{\fr {n \pi x} L}$,
and the eigenenergies are
$E_n = \fr {n^2 \pi^2} {2 L^2}$.
For simplicity,
we consider a box of length $L=1$ in atomic units.
With these values,
$T_{\rm cl} = \fr 2 {\bar n \pi}$ and
$t_{\rm rev} = \fr 4 \pi = 2\bar n T_{\rm cl}$
for a superposition centered on the value $\bar n$.
The discussion in Sec.\ II
shows that the wave packet is initially periodic
with period $T_{\rm cl}$
before subsequently collapsing and
exhibiting fractional and full revivals.
The full revivals are perfect revivals,
since the superrevival time $t_{\rm sr}$ is infinite.

Figure 2 shows the time evolution
of a wave packet in an infinite square well.
We have used
Eq.\ \rf{gauss}
for the $c_n$ coefficients with $\bar n = 15$ and $\si = 1.5$.
Figure 2a shows the initial wave packet at the left side of
the box near the origin.
It is highly oscillatory because it is
receiving an impulse from the infinite potential well.
Away from the walls,
Figure 2b,
it is smoother since it moves freely.
Figures 2c and 2d show the wave packet
colliding with the right-hand wall
and moving back to the left,
respectively.
At time $T_{\rm cl}$,
the wave packet has completed one cycle and is
shown in Fig.\ 2e.
Note that the shape has changed.
The wave packet is collapsing.
At time $t_{\rm rev}$,
however,
the wave packet regains its initial shape.
In fact,
the shape of the wave packet at any time $t$
exactly matches the shape at time
$t$ plus an integer multiple of $t_{\rm rev}$.

\vglue 0.6cm
{\bf\noindent C. Rigid Rotator}
\vglue 0.4cm

A quantum system closely related to the infinite
square well is the rigid rotator in two dimensions.
The hamiltonian is
$H = \fr {L_z^2} {2I}$,
where $L_z$ is the angular momentum
and $I$ is the moment of inertia.
For simplicity,
we set $I=1$ in atomic units.
The eigenfunctions obey periodic boundary conditions and
are given by $\psi_n(\ph) = \fr 1 {\sqrt{2 \pi}} e^{i n \ph}$,
and the energies are $E_n = n^2 /2$.
The analysis of revival structure in Sec.\ II
predicts the same behavior
for the rigid rotator as for the particle in
an infinite square well,
since both problems have quadratic energy dependence on $n$.

Figure 3 shows a wave packet for the rigid rotator with
$\bar n = 15$ and $\si = 1.5$.
Since $\bar n$ is positive,
the wave packet moves counterclockwise
(Figures 3a, 3b, 3c).
The motion is initially periodic,
but the shape of the wave packet changes as it moves.
For times greater than $T_{\rm cl}$,
the wave packet spreads and collapses.
Figure 3d shows a fractional revival.
At $t = \fr 1 4 t_{\rm rev}$,
the wave packet has reformed
into two macroscopically distinct packets moving with period
$\fr 1 2 T_{\rm cl}$.
Figure 3e shows another fractional revival at
$t = \fr 1 3 t_{\rm rev}$,
consisting of three distinct
subsidiary wave packets moving with periodicity
$\fr 1 3 T_{\rm cl}$.
At $t = \fr 1 2 t_{\rm rev}$,
a single wave packet forms (Figure 3f).
This wave packet resembles the initial wave packet,
but it is a half cycle out of phase with the classical motion.
The full revival at $t = t_{\rm rev}$ is a single wave packet
that exactly matches the initial one in Fig.\ 3a.

\vglue 0.6cm
{\bf\noindent D. Rydberg Wave Packets}
\vglue 0.4cm

Depending on the excitation method,
several different types of Rydberg wave packets can be considered,
each with different geometry for the orbital motion.
We consider a superposition of circular states having
$l=m=n-1$.
These states have maximal values for the expectations
$\vev{L_z}$ and $\vev{L^2}$ and produce a wave packet
tightly confined to the $x$-$y$ plane
and moving along a circular orbit
\cite{gaeta}.
We also restrict our analysis here to wave packets for hydrogen,
which have energies $E_n = -1/2n^2$.
In ref.\ \cite{sr2},
we examine the revival structure of Rydberg wave packets
for alkali-metal atoms using an analytical
supersymmetry-based quantum-defect model
\cite{sqdt}.
We find that the quantum defects cause additional
shifts in the time scales,
but the structure of the revivals is similar to
that given here for hydrogen.

We consider a superposition of circular hydrogenic
states $\psi_{nlm} (\vec r)$ with
$l = m = n-1$ in
Eq.\ \rf{expand}
and a gaussian superposition
\rf{gauss}
for the $c_n$ coefficients.
The time scales are given by
$T_{\rm cl} = 2 \pi {\bar n}^3$,
$t_{\rm rev} = \fr {2 \bar n} 3 T_{\rm cl}$,
and $t_{\rm sr} = \fr {3 \bar n} 4 t_{\rm rev}$
in atomic units.
The analysis of revival structure in Sec.\ II
shows that a Rydberg wave packet
of this type initially follows the classical
periodic motion of a particle in a circular orbit.
After several cycles,
the wave packet spreads and collapses.
At times that are rational fractions of $t_{\rm rev}$,
it exhibits fractional revivals.
A full revival occurs at time $t_{\rm rev}$.
Fractional superrevivals occur at times that are
rational fractions of $t_{\rm sr}$.
These culminate at time $\fr 1 6 t_{\rm sr}$ with the
formation of a full superrevival.

Figure 4 shows the time evolution of a circular wave packet
with $\bar n = 120$ and $\si = 2.5$.
Converting the time to nanoseconds,
we have
$T_{\rm cl} \simeq 0.26$ nsec,
$t_{\rm rev} \simeq 21.0$ nsec,
and $t_{\rm sr} \simeq 1890$ nsec.
Figure 4a shows the initial wave packet as a function
of position in the $x$-$y$ plane.
Throughout the motion,
it remains at a constant mean radial distance
$\vev{r} \approx \fr 1 2 \bar n (2 \bar n + 1)$ given
by the expectation value of $r$ in the state with $l=m=n-1$.
For $\bar n = 120$,
$\vev{r} \simeq 14460$ a.u.
When the wave packet
returns to its initial position at the end of the first orbit,
it has spread (Figures 4b, 4c).
It subsequently collapses and exhibits fractional revivals.
Figure 4d shows the fractional revival at
$\fr 1 4 t_{\rm rev}$
with two subsidiary wave packets.
As expected,
the full revival (Figure 4e) at time $t_{\rm rev}$
does not exactly match the initial wave packet.
The full superrevival at $\fr 1 6 t_{\rm sr}$ is
shown in Fig.\ 4f.
This wave packet resembles the initial wave packet better
than the full revival does at time $t_{\rm rev}$.

The example considered here uses a relatively large value
of $\bar n$ equal to 120 to illustrate the revival structure.
Experiments that have detected full and fractional revivals
of Rydberg wave packets have involved values in the range
$\bar n \simeq 40$ -- $90$,
corresponding to revival times in the range
$t_{\rm rev} \simeq 1$ -- $6$ nsec.
The apparatus in these experiments requires using a delay
line on the order of a few nanoseconds between pump and
probe signals to excite and then ionize the Rydberg atoms.
A full superrevival at time $\fr 1 6 t_{\rm sr}$ can be
detected with a similar delay line for wave packets with
$\bar n \simeq 25$ -- $50$.
For example,
with $\bar n = 40$,
the full superrevival occurs after approximately $1.3$ nsec,
making detection possible with current experimental apparatus.

\vglue 0.6cm
{\bf\noindent IV. AUTOCORRELATION}
\vglue 0.4cm

The autocorrelation function for a time-evolved
wave function with an initial wave function is given as
$A(t) = \vev{\Psi (\vec r,0) | \Psi (\vec r,t)}$.
Its absolute square gives a measure of the overlap
between the wave packet at time $t=0$ and at a later
time $t$.
Substituting the expansion in
Eq.\ \rf{expand}
into $|A(t)|^2$ yields
\beq
\vert A(t) \vert^2 = {\Bigm| \sum_n \vert c_n
\vert^2 e^{-i E_{n} t} \Bigm|}^2
\quad .
\label{auto}
\eeq
Using Eq.\ \rf{gauss} for the $c_n$ coefficients,
we can plot $|A(t)|^2$ directly for the different
quantum systems.

The autocorrelation function
gives a striking illustration of the different
types of revival structures.
Numerically,
$|A(t)|^2$ varies between 0 and 1.
When a wave packet exactly matches the initial wave packet,
$|A(t)|^2$ equals 1.
If the wave packet is far from its initial position,
then $|A(t)|^2 \simeq 0$.
Furthermore,
the periodicities in $|A(t)|^2$ reveal the periodic
behavior of the wave function.
In particular,
fractional revivals and fractional superrevivals
appear as periodic peaks in the autocorrelation function with
periods that are rational fractions of $T_{\rm cl}$ and
$t_{\rm rev}$,
respectively.

Figure 5 shows the absolute square of the autocorrelation function
as a function of time for four examples.
The rigid rotator is excluded because its
autocorrelation function is similar to that of
the particle in an infinite square well.

Figure 5a shows the autocorrelation for
the evolution of a free-particle wave packet in one dimension.
The wave packet consists of a gaussian-weighted
superposition of harmonic waves,
with momentum-dependent coefficients
\beq
| \ph (p) |^2 = \fr 1 {\sqrt{2 \pi \si^2}}
e^{\fr {-(p-p_0)^2} {2 \si^2}}
\quad .
\label{transform}
\eeq
We choose $\si = 2.5$ and $p_0 = 10$ in atomic units.
The wave packet starts at the origin and moves away,
spreading and gradually collapsing.
At $t=0$,
the autocorrelation is equal to 1.
For times $t>0$,
it decreases and asymptotically approaches 0.
There is no periodicity and no revivals.

Figure 5b shows the autocorrelation
for the harmonic oscillator.
The wave packet is readily seen to be perfectly
periodic with the period $T_{\rm cl} = 2 \pi$ in atomic units.
The autocorrelation oscillates with the wave packet.
The wave packet does not collapse,
and there are no revivals.

The autocorrelation for the particle in an infinite square
well is shown in Fig.\ 5c.
Here,
$T_{\rm cl} \simeq 0.04$ a.u.\ and
$t_{\rm rev} \simeq 1.27$ a.u.
The initial periodicity of the wave packet is revealed
by the periodicity in the peaks of $|A(t)|^2$ near $t=0$,
with period $T_{\rm cl}$.
The peaks decrease with time,
corresponding to the collapse of
the wave packet.
Fractional revivals are evident at
$\fr 1 4 t_{\rm rev} \simeq 0.32$ a.u.,
$\fr 1 2 t_{\rm rev} \simeq 0.64$ a.u.,
and $\fr 3 4 t_{\rm rev} \simeq 0.95$ a.u.
The periodicity of $|A(t)|^2$ at
$\fr 1 4 t_{\rm rev}$ and $\fr 3 4 t_{\rm rev}$
is $\fr 1 2 T_{\rm cl}$ and is evident in the graph in
that the peaks are much closer together at these times.
The autocorrelation $|A(t)|^2 = 1$ at $t_{\rm rev}$,
showing that the full revival is a perfect revival.
The periodicity near $t_{\rm rev}$ is equal to $T_{\rm cl}$
again.
This cycle of periodicity, collapse, and fractional/full
revivals repeats at multiples of $t_{\rm rev}$.
There are no superrevivals.

Figure 5d is the autocorrelation for the circular Rydberg
wave packets considered in Sec.\ III,
with the time given in nanoseconds.
The initial periodicity
$T_{\rm cl} \simeq 0.26$ nsec is difficult to resolve on the
scale used in this graph.
However,
the full revival at $t_{\rm rev} \simeq 21.0$ nsec as
well as some of the fractional revivals are evident.
The large peaks at
$t \approx \fr 1 2 t_{\rm rev} \simeq 10.5$ nsec
indicate the formation of a single wave packet.
This revival,
however,
is one half cycle out of phase with the classical motion.
The full revival at $t \approx t_{\rm rev}$ is
evidently not a perfect revival since $|A(t)|^2$ is
less than 1.
Fractional superrevivals appear at
$t \approx \fr 1 {18} t_{\rm sr} \simeq 105$ nsec
and $t \approx \fr 1 {12} t_{\rm sr} \simeq 158$ nsec
with periodicities $\fr 1 6 t_{\rm rev}$ and
$\fr 1 4 t_{\rm rev}$,
respectively.
The superrevival occurs at
$t \approx \fr 1 {6} t_{\rm sr} \simeq 315$ nsec,
and the periodicity is $\fr 1 2 t_{\rm rev}$.
The height of the peaks at
$t \approx \fr 1 6 t_{\rm sr}$ are greater than those
at $t_{\rm rev}$ indicating that the superrevival
resembles the initial wave packet better than does the
full revival at $t \approx t_{\rm rev}$.

\vglue 0.6cm
{\bf\noindent V. SUMMARY}
\vglue 0.2cm

In this paper,
a general treatment of the evolution and revival
structure of localized quantum wave packets is presented.
We consider wave packets formed
from a superposition of states heavily peaked
around a mean value $\bar n$ of the relevant quantum number,
as would occur in an excitation process using a
short-pulsed laser.
The gaussian form for the weighting coefficients in
Eq.\ \rf{gauss}
is used to simplify the calculations in the examples,
but the results do not depend on this form.
Three time scales
$T_{\rm cl}$, $t_{\rm rev}$, and $t_{\rm sr}$
controlling the evolution and revivals of wave packets
are introduced.
They are defined in Eq.\ \rf{Tcl}
and are governed only by the dependence of the energy
$E_n$ on the quantum numbers $n$.

We analyze the revival structure of
localized quantum wave packets for different quantum systems.
Given the form of the energy spectrum $E_n$ for a given
quantum system,
it is possible to predict the motion and types of
revivals that a wave packet will undergo.
We show that wave packets in
certain classes of quantum systems
exhibit particularly simple revival structures,
with perfect classical periodicity or perfect full revivals.
These results are illustrated
with several examples taken from quantum systems
that are commonly used in classes at the advanced undergraduate
or first-year graduate level
and that form the foundation for much of
the analysis of wave packets in the literature.
Plots are provided showing the wave packets and the
autocorrelation functions for these examples.
The revival structures analyzed here are
applicable to a multitude of physical systems
using wave packets in physics and chemistry.

\vfill
\eject

\vglue 0.6cm
{\bf\noindent ACKNOWLEDGMENTS}
\vglue 0.4cm

This work is supported in part by the National
Science Foundation under grant number PHY-9503756.

\baselineskip=20pt

\vglue 0.6cm
{\bf\noindent REFERENCES}
\vglue 0.4cm

\vfill\eject

\baselineskip=16pt

\begin{description}

\item[{\rm Fig.\ 1:}]
Unnormalized quantum wave packets for the simple
harmonic oscillator
with ${\bar n} = 15$ and $\si = 1.5$.
The probability density is plotted as
a function of $x$ in atomic units at the times:
(a) $t=0$ mod $T_{\rm cl}$,
(b) $t = \fr 1 4 T_{\rm cl}$ mod $T_{\rm cl}$,
(c) $t = \fr 1 2 T_{\rm cl}$ mod $T_{\rm cl}$,
(d) $t = \fr 3 4 T_{\rm cl}$ mod $T_{\rm cl}$.

\item[{\rm Fig.\ 2:}]
Unnormalized quantum wave packets for a particle
in an infinite square well
with ${\bar n} = 15$ and $\si = 1.5$.
The probability density is plotted as
a function of $x$ in atomic units at the times:
(a) $t=0$ mod $t_{\rm rev}$,
(b) $t = \fr 1 4 T_{\rm cl}$ mod $t_{\rm rev}$,
(c) $t = \fr 1 2 T_{\rm cl}$ mod $t_{\rm rev}$,
(d) $t = \fr 3 4 T_{\rm cl}$ mod $t_{\rm rev}$,
(e) $t = T_{\rm cl}$ mod $t_{\rm rev}$.

\item[{\rm Fig.\ 3:}]
Unnormalized quantum wave packets for the rigid rotator
with ${\bar l} = 15$ and $\si = 1.5$.
The probability density is plotted as
a function of $\ph$ in radians at the times:
(a) $t=0$ mod $t_{\rm rev}$,
(b) $t = \fr 1 2 T_{\rm cl}$ mod $t_{\rm rev}$,
(c) $t = T_{\rm cl}$ mod $t_{\rm rev}$,
(d) $t = \fr 1 4 t_{\rm rev}$ mod $t_{\rm rev}$,
(e) $t = \fr 1 3 t_{\rm rev}$ mod $t_{\rm rev}$,
(f) $t = \fr 1 2 t_{\rm rev}$ mod $t_{\rm rev}$.

\item[{\rm Fig.\ 4:}]
Unnormalized circular wave packets with ${\bar n} = 72$
and $\si = 2.5$.
The probability density is plotted as
a function of $x$ and $y$ in atomic units at the times:
(a) $t=0$,
(b) $t = \fr 1 2 T_{\rm cl}$,
(c) $t = T_{\rm cl}$,
(d) $t = \fr 1 4 t_{\rm rev}$,
(e) $t = t_{\rm rev}$,
(f) $t = \fr 1 6 t_{\rm sr}$.

\item[{\rm Fig.\ 5:}]
The absolute square of the autocorrelation function
as a function of time for four different quantum systems:
(a) the free particle,
(b) the simple harmonic oscillator,
(c) the infinite square well,
(d) Rydberg wave packets.

\end{description}

\vfill
\eject


\begin{thebibliography}{xx}

\bibitem{ps}
J. Parker and C.R. Stroud,
``Coherence and decay of Rydberg wave packets,''
Phys. Rev. Lett. {\bf 56}, 716-719 (1986);
``Rydberg wave packets and the classical limit,''
Phys. Scr. {\bf T12}, 70-75 (1986).

\bibitem{az}
G. Alber, H. Ritsch, and P. Zoller,
``Generation and detection of Rydberg wave packets by
short laser pulses,''
Phys. Rev. A {\bf 34}, 1058-1064 (1986);
G. Alber and P. Zoller,
``Laser Excitation of Electronic Wave Packets in
Rydberg Atoms,''
Phys. Rep. {\bf 199}, 231-280 (1991).

\bibitem{sciam}
M.\ Nauenberg, C.\ Stroud, and J.\ Yeazell,
``The classical limit of an atom,''
Sci.\ Am.\ {\bf 270}, 44-49 (June 1994).

\bibitem{tenWolde}
A. ten Wolde, L.D. Noordam, A. Lagendijk,
and H.B. van Linden van den Heuvell,
``Observation of radially localized atomic electron wave packets,''
Phys. Rev. Lett. {\bf 61}, 2099-2101 (1988).

\bibitem{yeazell1}
J.A. Yeazell, M. Mallalieu, J. Parker, and C.R. Stroud,
``Classical periodic motion of atomic-electron wave packets,''
Phys. Rev. A {\bf 40}, 5040-5043 (1989).

\bibitem{ap}
I.Sh. Averbukh and N.F. Perelman,
``Fractional revivals:  universality in the long-term evolution
of quantum wave packets beyond the correspondence principle dynamics,''
Phys. Lett. {\bf 139A}, 449-453 (1989).

\bibitem{nau1}
M. Nauenberg,
``Autocorrelation function and quantum recurrence of wavepackets,''
J. Phys. B {\bf 23}, L385-L390 (1990).

\bibitem{peres}
A. Peres,
``Multiple time scales for recurrences of Rydberg states,''
Phys.~Rev.~A {\bf 47}, 5196-5197 (1993).

\bibitem{detunings}
R. Bluhm and V.A. Kosteleck\'y,
``Quantum defects and the long-term behavior of radial
Rydberg wave packets,''
Phys. Rev. A {\bf 50}, R4445-R4448 (1994)
(hep-ph/9410325).

\bibitem{sr}
R. Bluhm and V.A. Kosteleck\'y,
``Long-term evolution and revival structure of Rydberg
wave packets,''
Phys.\ Lett.\ {200A}, 308-313 (1995)
(quant-ph/9508024).

\bibitem{sr2}
R. Bluhm and V.A. Kosteleck\'y,
``Long-term evolution and revival structure of Rydberg wave
packets for hydrogen and alkali-metal atoms,''
Phys.\ Rev.\ A {51}, 4767-4786 (1995)
(quant-ph/9506009).

\bibitem{bocch}
P.\ Bocchieri and A.\ Loinger,
``Quantum recurrence theorem,''
Phys.~Rev.\ {\bf 107}, 337-338 (1957).

\bibitem{almost}
H.A.\ Bohr,
\it Almost Periodic Functions, \rm
(Chelsea, New York, 1947).

\bibitem{yeazell2}
J.A. Yeazell, M. Mallalieu, and C.R. Stroud,
``Observation of the collapse and revival of a Rydberg
electronic wave packet,''
Phys. Rev. Lett. {\bf 64}, 2007-2010 (1990).

\bibitem{yeazell3}
J.A. Yeazell and C.R. Stroud,
``Observation of fractional revivals in the evolution of a
Rydberg atomic wave packet,''
Phys. Rev. A {\bf 43}, 5153-5156 (1991).

\bibitem{meacher}
D.R. Meacher, P.E. Meyler, I.G. Hughes, and P. Ewart,
``Observation of the collapse and revival of a Rydberg
wavepacket in atomic rubidium,''
J. Phys. B {\bf 24}, L63-L69 (1991).

\bibitem{wals}
J. Wals, H.H. Fielding, J.F. Christian, L.C. Snoek,
W.J. van der Zande, and H.B. van Linden van den Heuvell,
``Observation of Rydberg wave packet dynamics in a coulombic
and magnetic field,''
Phys.\ Rev.\ Lett.\ {\bf 72}, 3783-3786 (1994).

\bibitem{garraway}
For a review and additional references, see
B.M.\ Garraway and K-A. Suominen,
``Wave-packet dynamics: new physics and chemistry in
femto-time,''
Rep.~Prog.~Phys.\ {\bf 58}, 365-419 (1995).

\bibitem{tannor}
D.J.\ Tannor and S.A.\ Rice,
``Control of selectivity of chemical reaction via control
of wave packet evolution,''
J.~Chem.~Phys.\ {\bf 83}, 5013-5018 (1985).

\bibitem{warren}
W.S.\ Warren, H.\ Rabitz, M.\ Dahleh,
``Coherent control of quantum dynamics: the dream is alive,''
Science {\bf 259}, 1581-1589 (1993).

\bibitem{wilson}
B.\ Kohler, V.V.\ Yakovlev, J.\ Che, J.L.\ Krause,
M.\ Messina, and K.R.\ Wilson,
``Quantum control of wave packet evolution with tailored
femtosecond pulses,''
Phys.~Rev.~Lett.\ {\bf 74}, 3360-3363 (1995).

\bibitem{leo}
K.\ Leo, J.\ Shah, E.O.\ Gobel, and T.C.\ Damen,
``Coherent oscillations of a wave packet in a semiconductor
double-quantum-well structure,''
Phys.~Rev.~Lett.\ {\bf 66}, 201-204 (1991).

\bibitem{gobel}
E.O.\ Gobel, K.\ Leo, T.C.\ Damen, and J.\ Shah,
``Quantum beats of excitons in quantum wells,''
Phys.~Rev.~Lett.\ {\bf 64}, 1801-1804 (1990).

\bibitem{bs}
M.L.\ Biermann and C.R.\ Stroud,
``Wave-packet theory of coherent carrier dynamics in a
semiconductor superlattice,''
Phys.~Rev.~B {\bf 47}, 3718-3727 (1993).

\bibitem{rwk}
G.\ Rempe, H.\ Walther, and N.\ Klein,
``Observation of quantum collapse and revival in a
one-atom maser,''
Phys.~Rev.~Lett.\ {\bf 58}, 353-356 (1987).

\bibitem{avjc}
I.Sh. Averbukh,
``Fractional revivals in the Jaynes Cummings model,''
Phys.~Rev.~A {\bf 46}, R2205-R2208 (1992).

\bibitem{vet}
S.I.\ Vetchinkin and V.V.\ Eryomin,
``The structure of wavepacket fractional revivals in a
Morse-like anharmonic system,''
Chem.~Phys.\ Lett.\ {\bf 222}, 394-398 (1994).

\bibitem{cm}
W.-Y.\ Chen and G.J.\ Milburn,
``Fractional quantum revivals in the atomic gravitational cavity,''
Phys.~Rev.~A {\bf 51}, 2328-2333 (1995).

\bibitem{goldberg}
A.\ Goldberg, H.M.\ Schey, J.L.\ Schwartz,
``Computer-generated motion pictures of one-dimensional quantum
mechanical transmission and reflection phenomena,''
Am.\ J.\ Phys.\ {\bf 35}, 177-186 (1967).

\bibitem{picture}
S.\ Brandt and H.D.\ Dahmen,
\it The Picture Book of Quantum Mechanics \rm,
2nd ed.,
(Springer-Verlag, New York, 1995).

\bibitem{klauder}
See, for example,
J.R. Klauder and B.-S. Skagerstam, eds.,
{\it Coherent States}
(World Scientific, Singapore, 1985).

\bibitem{nieto}
M.M. Nieto,
``Coherent states for general potentials IV: three-dimensional
systems,''
Phys. Rev. D {\bf 22}, 391-402 (1980);
V.P. Gutschick and M.M. Nieto,
``Coherent states for general potentials V: time evolution,''
Phys. Rev. D {\bf 22}, 403-418 (1980).

\bibitem{rss}
R. Bluhm and V.A. Kosteleck\'y,
``Radial squeezed states and Rydberg wave packets,''
Phys. Rev. A {\bf 48}, R4047-R4050 (1993)
(quant-ph/9508019).

\bibitem{susywave}
R. Bluhm and V.A. Kosteleck\'y,
``Atomic supersymmetry, Rydberg wave packets,
and radial squeezed states,''
Phys. Rev. A {\bf 49}, 4628-4640 (1994)
(quant-ph/9508020).

\bibitem{primers}
For primers on coherent states and squeezed states, see
H.A.\ Gersch,
``Time evolution of minimum uncertainty states of a
harmonic oscillator,''
Am.\ J.\ Phys.\ {\bf 60}, 1024-1030 (1992);
R.W.\ Henry and S.C.\ Glotzer,
``A squeezed-state primer,''
Am.\ J.\ Phys.\ {\bf 56}, 318-328 (1988);
M.M. Nieto, in
G.T. Moore and M.O. Scully, eds.,
{\it Frontiers of Nonequilibrium Statistical Physics}
(Plenum, New York, 1986), pp. 287-307.

\bibitem{gns}
Z.D. Gaeta, M. Noel, and C.R. Stroud,
``Excitation of the classical-limit state of an atom,''
Phys. Rev. Lett. {\bf 73}, 636-639 (1994).

\bibitem{2dss}
R. Bluhm, V.A. Kosteleck\'y, and B. Tudose,
``Elliptical Squeezed States and Rydberg Wave Packets,''
Phys. Rev. A {\bf 52}, 2234-2244 (1990)
(quant-ph/9509010);
``Keplerian Squeezed States and Rydberg Wave Packets,''
Phys. Rev. A, in press
(quant-ph/9510023).

\bibitem{ns}
M.M.\ Nieto and L.M.\ Simmons,
``Limiting spectra from confining potentials,''
Am.\ J.\ Phys.\ {\bf 47}, 634-635 (1979).

\bibitem{gaeta}
Z. Da{\v c}i\'c Gaeta and C.R. Stroud,
``Classical and quantum-mechanical dynamics of a quasiclassical
state of the hydrogen atom,''
Phys. Rev. A {\bf 42}, 6308-6313 (1990).

\bibitem{sqdt}
V.~A.~Kosteleck\'y and M.~M.~Nieto,
``Analytical wave functions for atomic quantum-defect theory,''
Phys.~Rev.~A {\bf 32}, 3243-3246 (1985).
The basic idea of atomic supersymmetry is presented in
V.~A.~Kosteleck\'y and M.~M.~Nieto,
``Evidence for a phenomenological supersymmetry in atomic physics,''
Phys.~Rev.~Lett.~{\bf 53}, 2285-2288 (1984).
For a review, see, for example,
V.A. Kosteleck\'y,
``Atomic Supersymmetry,''
in B. Gruber and T. Osaka, eds.,
\it Symmetries in Science VII:
Dynamic Symmetries and Spectrum-Generating Algebras
in Physics, \rm
Plenum, New York, 1993
(quant-ph/9508015).

\end{thebibliography}
\end{document}